\documentclass{optica-article}

\journal{opticajournal} 

\articletype{Research Article}

\usepackage[separate-uncertainty=true]{siunitx}

\begin{document}

\title{Longitudinal to transversal conversion of mode-\\locked states in an empty optical resonator}

\author{Michael Zwilich,\authormark{*} Florian Schepers, and Carsten Fallnich}

\address{University of Münster, Institute of Applied Physics, Corrensstr. 2, 48149 Münster, Germany}

\email{\authormark{*}m.zwilich@uni-muenster.de} 


\begin{abstract*} 
A longitudinal mode-locked state can be converted to a transverse mode-locked state by exploiting the spectral and spatial filtering of an empty optical resonator. 
Carrier and amplitude modulation sidebands were simultaneously transmitted by the conversion resonator, yielding phase-locked superpositions of up to five transverse modes.
Equivalently, an amplitude-modulated beam was converted into a beam that periodically moved across the transverse plane.
Precise control over the spatial beam shape during oscillation was gained by independently altering the set of transverse modes and their respective powers, which demonstrated an increased level of control in the generation of transverse mode-locked states.
\end{abstract*}

\section{Introduction}
Optical resonators are routinely used as optical spectrum analyzers, where spectral components of an incident beam are distinguished because of the difference in phase acquired upon propagation~\cite{Fork1964, Hercher1968}.
Resonators are also utilized as mode-cleaners, where the spatial profile of a beam is decomposed into its transverse eigenmodes, which can in turn be differentiated because of their Gouy phase shifts~\cite{Siegman1986, BayerHelms1984, Sayeh1985}.
These spectral and spatial filtering properties make optical resonators a widely applied tool for measuring or stabilizing emission frequencies of light sources~\cite{Drever1983} and controlling the spatial structure of beams~\cite{Willke1998, Gossler2003, Kwee2007}, for example in the context of interferometrical gravitational wave detectors.
In these applications care is taken to transmit the individual, i.e., spectral and/or spatial, beam components separately, so that they can be resolved or filtered. 

In contrast, to simultaneously transmit multiple spectral components, e.g., to pass carrier and phase modulation sidebands through a mode-cleaner, the incident frequency spacing can be chosen to match the longitudinal mode spacing of the resonator~\cite{Skeldon1997}.
Similarly, carrier and sidebands can also be transmitted simultaneously by matching the modulation frequency to the transverse mode spacing, resulting in a superposition of neighboring transverse modes if the incident beam is not properly mode-matched~\cite{Anderson1984, Sampas1990}.
The so transmitted modes exhibit spatio-temporal beating due to their different spatial structures and resonance frequencies, which can be used as a control signal to align incident beam and resonator in order to actively suppress higher-order transverse modes.
In contrast, transverse modes and their superpositions are deliberately generated in the research field of transverse mode-locking due to the emerging spatio-temporal dynamics~\cite{Auston1968a}, that depend on the excited transverse mode family~\cite{Auston1968, Haug1974, Pariente2015, Schepers2020, Schepers2021}. 
A mode-locked state of Hermite-Gaussian modes, for example, can result in a Gaussian spot that periodically scans the transverse plane, with spot resolving rates in the GHz range already demonstrated~\cite{Schepers2020}.
Commonly, transverse mode-locked states are generated in laser-active resonators, where loss~\cite{Auston1968, Dukhovnyi1971, Schepers2020} and gain~\cite{Dukhovnyi1971, Liang2013, Shen2018, Schepers2020} are shaped according to the different spatial intensity distributions of the transverse modes that are intended to be excited. 
However, intensity distributions of transverse modes are not mutually exclusive, as they have spatial overlap with each other and even become more similar with increasing mode order.
Therefore, the level of control over the set of locked modes and their powers is limited when only exploiting the transverse modes' spatial features.

In this work, both the spatial and spectral filtering properties of an optical resonator are leveraged in the context of transverse mode-locking.
Transverse mode-locked states are generated via conversion of incident longitudinal mode-locked states by a simple, empty optical resonator.
In the frequency domain this is understood as matching the incident spectral components to resonance frequencies of the resonator's transverse modes.
As a result, in the time domain an amplitude-modulated beam is converted into a beam that periodically moves across the transverse plane, thus converting a longitudinal to a transverse mode-locked state.
Multiple mode-locked states with up to five transverse modes were successfully obtained by independently controlling the amplitudes of the incident spectral components, the spatial overlap with higher-order transverse modes, and the central mode order.
In this conversion scheme the transverse modes are not only addressed by their increasingly similar spatial intensity distributions but additionally by their distinct resonance frequencies, thus increasing the level of control in the generation of transverse mode-locked beams.

\section{Resonator eigenmodes and modal conversion}
In the following a simple optical two-mirror resonator is considered whose eigenmodes shall be well described by the Hermite-Gaussian (HG) mode set. 
The resonance frequencies $\nu_{qmn}$ of such a Gaussian resonator are given by~\cite{Siegman1986}
\begin{equation}
	\nu_{qmn} = q\Delta\nu_\text{L} + (m+n+1)\Delta\nu_\text{T}.
	\label{eq:resonance_frequencies}
\end{equation}
When only considering plane waves, the resonance frequencies are an integer multiple $q$ of the longitudinal mode spacing $\Delta\nu_\text{L} = c/2L$ defined by the resonator length $L$ and the speed of light $c$. 
When also taking the transverse shape of the electric field into account, one finds that the resonance frequencies are shifted linearly with increasing transverse mode orders $m$ and $n$ by the transverse mode spacing $\Delta\nu_\text{T}=\Delta\nu_\text{L} \psi/\pi$, with $\psi$ being the Gouy phase shift experienced by the fundamental mode ($m=n=0$) after a single pass through the resonator~\cite{Siegman1986}.

If the frequency $\nu$ of a single-frequency laser beam incident on a resonator matches the resonance frequency $\nu_{qmn}$ of a certain mode, the beam is transmitted. 
The transmitted power, however, also depends on the spatial overlap of the incident field $E_0(x,y)$ with the corresponding resonator mode $\text{HG}_{mn} (x,y)$, which is given by the absolute square of the coupling coefficient~\cite{BayerHelms1984}
\begin{equation}
	B_{mn} = |b_{mn}|^2 = \left|\iint_{-\infty}^{\infty} E_0(x,y) \text{HG}^*_{mn} (x,y) \,\mathrm{d}x\, \mathrm{d}y\right|^2.
	\label{eq:coupling_coefficient}
\end{equation}
The remaining, spectrally and spatially nonresonant beam content is reflected.
Therefore, it depends not only on the frequency of an incident wave but also on its transverse shape, whether it is resonant inside a resonator and can be transmitted to a substantial amount.
Throughout this section, $x$ and $y$ are the transverse spatial coordinates orthogonal to the resonator axis and $t$ represents time. 

\begin{figure}[h!]
	\centering
	\includegraphics{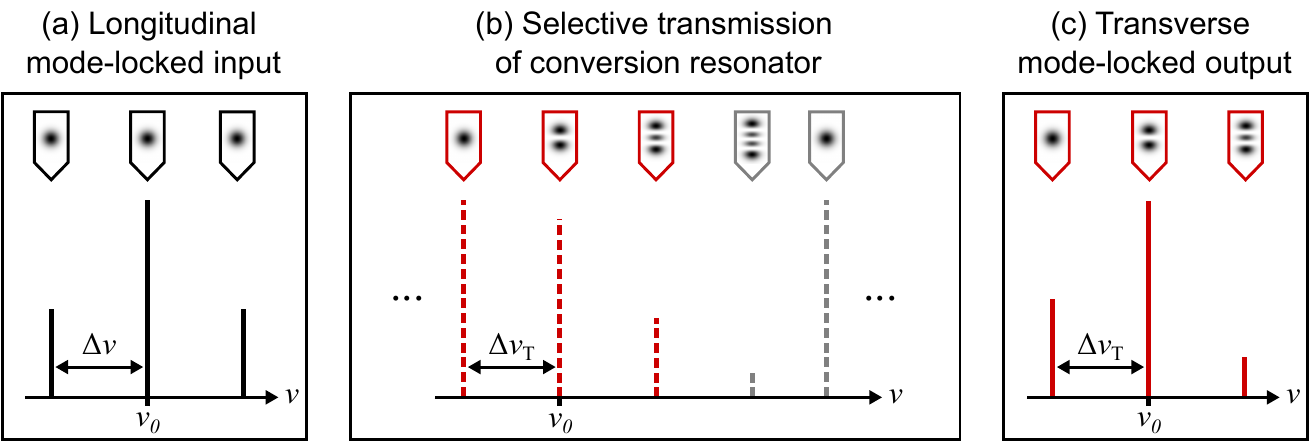}
	\caption{Schematic of the mode conversion scheme. (a) Multiple incident spectral components can be (b) transmitted by an empty optical resonator if their frequencies match transverse mode resonance frequencies (depicted as dashed red lines), resulting in (c) a superposition of transverse modes. If the incident spectral components are phase-locked, so are the transmitted transverse modes, resulting in longitudinal to transverse mode-locked conversion.}
	\label{fig:scheme}
\end{figure}

This reasoning can be extended to a beam consisting of multiple spectral components, which is shown schematically in Fig.~\ref{fig:scheme}(a) for three spectral components. 
Assuming a beam incident on the optical resonator that is of the form
\begin{equation}
	E_\text{incident}(x,y,t) = E_0(x,y) \sum_{k=-K}^{K} a_k \exp(i 2 \pi t(\nu_0 + k\Delta\nu)) \exp(i\phi_k),
	\label{eq:long_mode_locked}
\end{equation}
with $2K+1$ spectral components with different field amplitudes $a_k$, phases $\phi_k$, and frequencies $\nu_0 + k\Delta\nu$ regularly spaced around a central frequency $\nu_0$, but a common transverse beam shape $E_0(x,y)$. 
All of these spectral components can be transmitted simultaneously, if they each match one of the resonance frequencies (see Fig.~\ref{fig:scheme}(b)) of the resonator. 
This is fulfilled, if the input frequency spacing $\Delta \nu$ is equal to the resonator's transverse mode spacing $\Delta\nu_\text{T}$. 
Then, all input frequencies can line up with resonance frequencies of neighboring transverse modes, yielding the transmitted field
\begin{equation}
	E_\text{transmitted}(x,y,t) = \sum_{k=-K}^{K} a_k b_{m_0+k} \text{HG}_{m_0+k}(x,y) \exp(i 2 \pi t(\nu_0 + k \Delta\nu_\text{T})) \exp(i\phi_k),
	\label{eq:transv_mode_locked}
\end{equation}
where only the excitation of $\text{HG}_{m,0}$ modes is considered and the second mode index is omitted for conciseness.
The transmitted beam still has multiple spectral components, but a different beam shape associated with each one: it is a superposition of adjacent transverse modes $\text{HG}_{m_0+k}$ (see Fig.~\ref{fig:scheme}(c)). 
In contrast, the incident spectral components have different frequencies and a common beam shape; they can therefore be regarded as longitudinal modes originating from another resonator.
If these incident longitudinal modes are phase-locked to each other, so are the transverse modes in the resulting beam.
Therefore, this simple optical resonator converts a longitudinal mode-locked input to a transverse mode-locked output.
The characteristic feature of such a mode-locked superposition of Hermite-Gaussian modes is that the beam performs a scanning motion in the transverse plane with a frequency equal to the frequency difference $\Delta \nu_\text{T}$ between the individual modes~\cite{Auston1968a}.
The conversion by the resonator thus transfers an oscillation from the temporal domain (longitudinal modes) to the spatio-temporal domain (transverse modes).

The spatio-temporal oscillation depends on the set of superposed transverse modes and their respective powers, i.e., the modal power distribution. 
The set of transverse modes can be characterized by the central mode order $m_0$, that is resonant for the carrier frequency $\nu_0$. 
This central mode order is determined by the resonator length modulo the incident beam's wavelength $\lambda=c/\nu_0$, such that the set of locked modes can be altered by tuning the resonator length by a fraction of a wavelength.
The field amplitude of each mode depends on the initial field amplitude $a_k$ as well as the spatial overlap between the incident beam shape $E_0(x,y)$ and the corresponding resonant transverse mode $\text{HG}_{m_0+k}(x,y)$ as given by the coupling coefficient $b_{m_0+k}$. 
Therefore, the power of each transverse mode is $W_{m_0+k} = |a_k b_{m_0+k}|^2 = A_k B_{m_0+k}$.
The sum of all modal powers gives the overall transmitted power, yielding a conversion efficiency of $\eta = \sum_{k=-K}^{K} A_k B_{m_0 +k} / \sum_{k=-K}^{K} A_k$, when compared to the incident beam's power.
To give an example, assuming equal amplitudes $A_k$, the conversion efficiency is at most the inverse of the number of incident spectral components, i.e., $\eta_\text{max}=\num{0.33}$ and \num{0.2} for three and five incident spectral components, respectively.

\section{Experimental setup}
\begin{figure}[h!]
	\centering
	\includegraphics{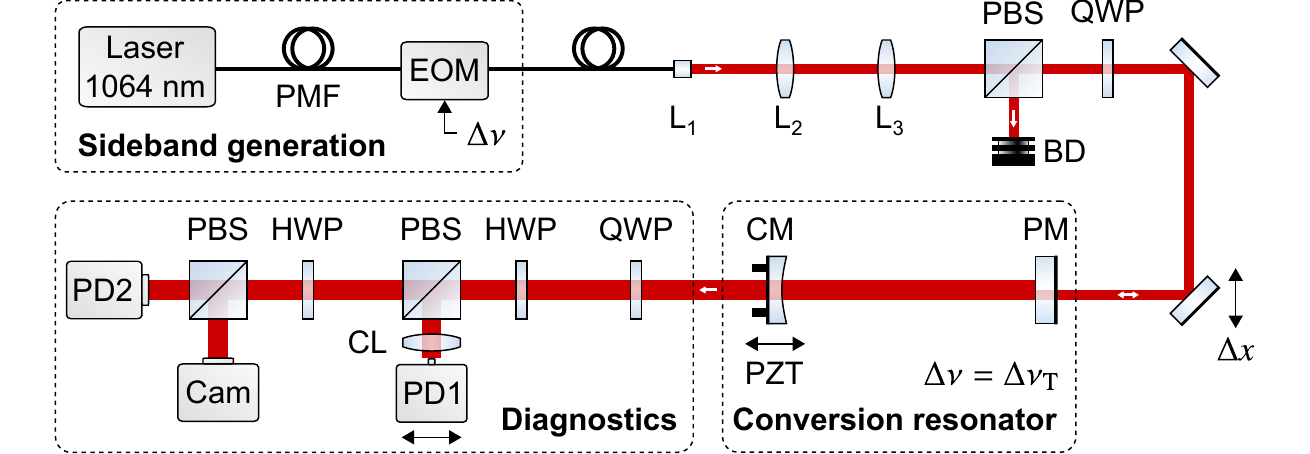}
	\caption{Schematic of the experimental setup. The modulation frequency $\Delta \nu$ matches the transverse mode-spacing $\Delta\nu_\text{T}$ of the resonator to allow simultaneous transmission of carrier and sidebands. The beam is displaced by $\Delta x$ from the resonator's optical axis to establish spatial overlap with higher-order transverse modes, which are indicated by the larger beam size in and after the conversion resonator. For details, see text.}
	\label{fig:setup}
\end{figure}

For our experimental realization (see Fig.~\ref{fig:setup}) of the proposed conversion scheme a continuous-wave monolithic nonplanar ring oscillator (wavelength $\lambda=\SI{1064}{\nm}$) was used as beam source because of its intrinsic stability as well as its longitudinal and transverse single-mode operation~\cite{Kane1985}.
To create a basic longitudinal mode-locked state, additional phase-locked spectral components were generated by amplitude modulation with a fiber-coupled (PMF, polarization maintaining fiber) lithium niobate-based Mach-Zehnder type electro-optic modulator (EOM, ixBlue NIR-MX-LN-10). 
Its low half-wave voltage of $V_\pi=\SI{4.3}{\V}$ allowed using a waveform generator (Siglent SDG6022X) to apply the RF control signal even at modulation frequencies $\Delta\nu$ of up to \SI{160}{\MHz}. 
By tuning the applied modulation and bias signals, the ratio of carrier and sideband amplitudes could be chosen freely, ranging from a state with zero amplitude sidebands to a state with zero amplitude carrier.

In order to create defined spatial overlap between the incident beam and higher-order transverse modes, the beam was first mode-matched to the fundamental mode of the resonator and then deliberately displaced.
This approach strikes a balance between versatility, i.e., realizing different transverse mode superpositions, and efficiency, i.e., the total transmitted power.
The beam was mode-matched with a lens arrangement ($\text{L}_1, \text{L}_2, \text{L}_3$), such that the measured beam size of \SI{412(10)}{\um} at the first resonator mirror matched the beam size $w_0=\SI{409}{\um}$ of the resonator's fundamental mode.
The horizontal displacement $\Delta x$ between incident beam and optical axis of the resonator was then introduced by moving the final steering mirror on a translation stage.
Thus, only transverse modes of the form $\text{HG}_{m,0}$ had spatial overlap with the incident beam and could be excited, with an arbitrary mode order $m$ in the horizontal direction and the fundamental mode order $n=0$ in the vertical direction.

A polarizing beamsplitter (PBS) and a quarter-wave plate (QWP) acting as an optically isolating element were placed in front of the resonator with a beam dump (BD), which reduced back reflections by \SI{30}{\dB}.
The resulting circular polarization was compensated by an inversely oriented quarter-wave plate behind the resonator.

The conversion resonator was formed by a plane mirror (PM) and a concave end-mirror (CM, radius of curvature $R=\SI{-1}{\m}$). 
The mirrors were each dielectrically coated with an anti-reflection coating on the outward and reflective coating on the inward facing side for a wavelength of \SI{1064}{\nm}, resulting in mirror power transmissions of \SI{2.1}{\percent} and \SI{1.0}{\percent}, respectively. 
The transverse mode spacing was $\Delta\nu_\text{T} = \SI{79.6(1)}{\MHz}$. 
From the longitudinal mode spacing of \SI{354.9(1)}{\MHz} a resonator length of \SI{422.3(1)}{\mm} was deduced, which could be varied by three piezoelectric actuators (PZT) spring-mounted to the back of the end-mirror.
Assuming a fixed carrier frequency, the excitation of a set of transverse modes requires a specific resonator length to allow the conversion and the according transmission of the incident longitudinal modes. 
For the following proof-of-principle experiments it was sufficient to scan the resonator length periodically over the range of a few wavelengths to ensure resonance of the incident beam's spectral components at recurring points in time.
At these instants the incident phase-locked spectral components around the carrier $\nu_0$ were resonant with a set of transverse modes around a central mode order $m_0$ and therefore being transmitted, resulting in a transverse mode-locked output beam. 
In future experimental realizations the carrier frequency could be stabilized with respect to a transverse mode resonance using a feedback loop~\cite{Liu2019} to maintain the mode-locked state over longer periods of time.

The spatio-temporal oscillation exhibited by the transmitted transverse mode-locked beam was spatially sampled by a photodetector (PD1, Femto HSPR-X-I-2G-IN), which was gradually moved along the beam's principal axis by a linear translation stage.
To increase signal levels, the beam was focused along its vertical direction using a cylindrical lens (CL), thereby preserving the beam shapes of the $\text{HG}_{m,0}$ modes along the horizontal direction.
It was made sure that the photodetector's aperture of \SI{100}{\um} was smaller than the transverse dimensions of the output beam with a fundamental beam size of \SI{904(10)}{\um} in the measurement plane, and the photodetector's bandwidth of \SI{2}{\GHz} was larger than the oscillation frequency of \SI{79.6}{\MHz} to achieve sufficient spatial as well as temporal resolution.
To measure the power $W_m$ of each transverse mode in a certain superposition, the incident beam was modulated with a frequency detuned from the transverse mode spacing after performing the mode-locking experiment.
The individual spectral and spatial beam components were then no longer transmitted simultaneously, but instead were resolved temporally while scanning the resonator length, so that carrier and sideband amplitudes $A_k$ as well as the spatial coupling coefficients $B_m$ could be determined.
This temporally resolved signal was detected by a photodiode (PD2, Hamamatsu S3590, amplified with up to \SI{100}{\kHz} bandwidth) with a $\num{10}\times\num{10}$~\si{\mm^2} large sensing area to avoid spatial clipping for higher-order transverse modes.
Additionally, the spatial intensity distributions were evaluated with a CCD camera (Cam) to ensure horizontal mode orientation. 
In order to set appropriate signal levels, the power incident on each detector could be varied with half-wave plates (HWP) and polarizing beamsplitters.

\section{Conversion of longitudinal to transverse mode-locked states}
\subsection{First experimental demonstration}
\begin{figure}[h!]
	\centering
	\includegraphics{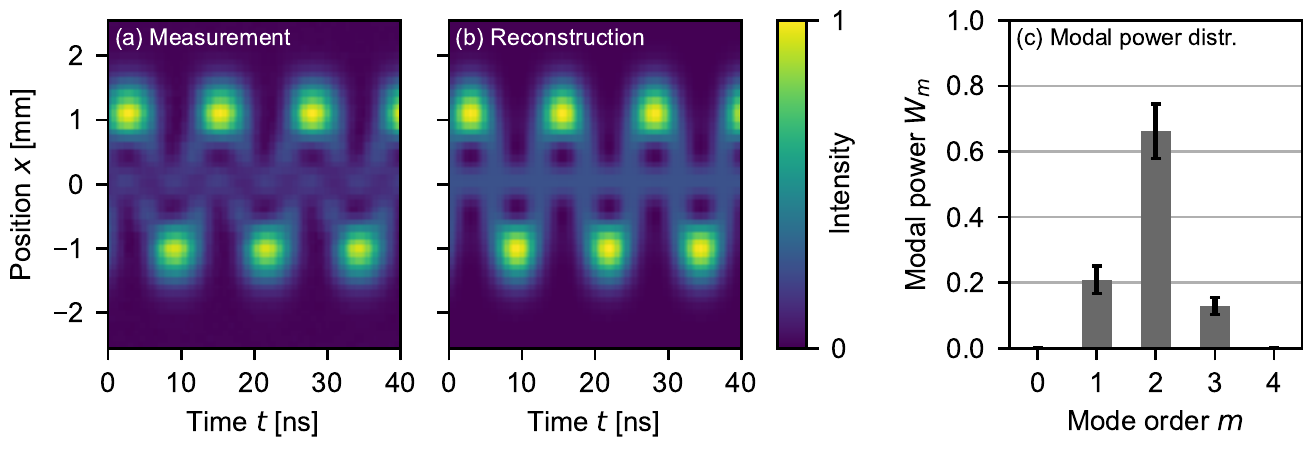}
	\caption{Transverse mode-locked state obtained by conversion in the resonator. (a) Measured spatio-temporal oscillation. (b) Reconstruction based on the measured modal power distribution shown in (c). Intensity and modal powers are given in arbitrary units normalized to maximum intensity and to their overall sum, respectively.}
	\label{fig:result1}
\end{figure}

With the presented conversion scheme, superpositions of transverse modes with different modal power distributions have been realized, depending on the amplitudes of the incident spectral components and the spatial overlap of the incident beam with the transverse resonator modes. 
As an example, a basic longitudinal mode-locked state consisting of three phase-locked spectral components was generated by amplitude modulation and converted using the empty optical resonator, whose transverse mode spacing $\Delta \nu_\text{T}$ matched the modulation frequency $\Delta \nu$. 
In front of the resonator a modulation of the beam's intensity at $\Delta \nu = \SI{79.6}{\MHz}$ was observed, whereas the resulting resonator transmission showed a spatio-temporal oscillation, where the beam's center periodically changed its position (see Fig.~\ref{fig:result1}(a)).
This oscillation is characteristic for transverse mode-locking and took place with a frequency equal to the transverse mode spacing of the resonator and modulation frequency of $\Delta \nu_\text{T} = \Delta \nu = \SI{79.6}{\MHz}$.
As can be seen, this simple optical setup converted an amplitude-modulated beam into a spatially oscillating beam, transferring an oscillation from the temporal to the spatio-temporal domain.

In order to quantitatively verify the presented conversion scheme, the expected spatio-temporal trajectory of each transverse mode-locked state was reconstructed numerically according to Eq.~\ref{eq:transv_mode_locked}.
In the following, the contributing factors are pointed out for the particular superposition shown in Fig.~\ref{fig:result1}.
The incident longitudinal mode-locked state was generated via amplitude modulation.
By selecting an appropriate modulation depth and bias point, the first order sideband amplitudes were chosen to be $A_{-1}=A_1=\num{0.30(06)}$ relative to that of the carrier ($A_0=\num{1.00(11)}$), making a compromise between sideband amplitude and overall transmitted power. 
In this example, only those instants -- while scanning the resonator length -- were considered, when the carrier frequency $\nu_0$ matched the resonance of the central mode order $m_0=2$.
To establish spatial overlap with higher-order transverse modes, the incident beam was displaced horizontally by $\Delta x = \SI{580}{\um}$ (roughly \num{1.2} times the fundamental beam size) relative to the optical axis of the resonator, which led to coupling coefficients $B_1=\num{0.27(02)}$, $B_2=\num{0.26(02)}$, and $B_3=\num{0.17(01)}$ for the relevant transverse modes around $m_0=2$.
Therefore, the resulting mode-locked state was expected to be a superposition of the $\text{HG}_{1,0}$ mode, excited by the lower sideband, the $\text{HG}_{2,0}$ mode, excited by the carrier, and the $\text{HG}_{3,0}$ mode, excited by the upper sideband, with their powers given by the product of initial spectral amplitudes $A_k$ and coupling coefficients $B_{m_0+k}$, yielding relative modal powers of $W_1=\SI{21(4)}{\percent}$, $W_2=\SI{66(8)}{\percent}$, and $W_3=\SI{13(3)}{\percent}$.
The uncertainties arise from fluctuations of the relative carrier and sideband amplitudes over time and from how precisely $A_k$ and $B_m$ could be determined from the photodetector (PD2) signal.
Based on these modal powers and the resonator's Hermite-Gaussian mode set the expected spatio-temporal trajectory for the superposition of the three transverse modes was computed (see Fig.~\ref{fig:result1}(b)). 
This reconstruction replicated the main features of the spatio-temporal oscillation, i.e., the periodic change of the beam's center position and the increased intensity at the turning points during the periodic evolution of the beam shape.
As a measure of similarity between the measured spatio-temporal trajectory $I(t,x)$ and its reconstruction $I'(t,x)$ the correlation coefficient
\begin{equation}
	\gamma = \frac{\sum \left(I\left(t,x\right)-\bar{I}\right) \left(I'\left(t,x\right)-\bar{I'}\right)}{\sqrt{\sum\left(I\left(t,x\right)-\bar{I}\right)^2 \sum\left(I'\left(t,x\right)-\bar{I'}\right)^2}}
\end{equation}
was calculated, where $\bar{I}$ and $\bar{I'}$ are the respective mean values.
During our experiments, and similar to previous studies~\cite{Schepers2020}, the key characteristics of the measured spatio-temporal trajectories were reproduced by their respective numerical reconstructions for $\gamma > 0.95$.
Compared to this benchmark, the correlation coefficient of $\gamma = \num{0.981}$ depicted in Fig.~\ref{fig:result1} indicated good agreement between measurement and reconstruction. 
This high degree of similarity further verified the presented conversion scheme and demonstrated the successful conversion of a longitudinal to a transverse mode-locked state.

Still, the measured and numerically reconstructed spatio-temporal trajectories differed in the center of the oscillation (around $x=\SI{0}{\mm}$).
At this position the intensities of the odd modes $\text{HG}_{1,0}$ and $\text{HG}_{3,0}$ are zero, which should only leave the even $\text{HG}_{2,0}$ mode with its central lobe, resulting in a constant intensity as seen in the reconstruction.
The measured spatio-temporal trajectory, however, showed only half of the expected constant intensity, likely because the used photodetector (PD1) was ac-coupled, leading to a decrease in amplitude of such slowly varying signals.
Additionally, due to the photodetector's finite spatial aperture residual interference between the odd modes with their frequency difference $2 \Delta \nu_\text{T}$ could have been picked up, resulting in the observed intensity modulation in the spatial center of the oscillation.

These measurement artifacts were not accounted for in the reconstruction, because the key features of the spatio-temporal oscillation were already reconstructed accurately as indicated by the high correlation coefficient $\gamma=0.981$. 
The good agreement between measured and reconstructed spatio-temporal trajectories is also evident when comparing other transverse mode-locked states and their respective reconstructions, as is done in comprehensive listings in Supplement~1.

\subsection{Controlling the modal power distribution}
A transverse mode-locked state is characterized by the set of locked modes and their respective powers (see Eq.~\ref{eq:transv_mode_locked}).
Here, to generate different transverse mode-locked states, the resonator length modulo the incident beam's wavelength, the displacement of the incident beam, and the amplitude modulation control signal were altered individually, compared to the superposition shown in Fig.~\ref{fig:result1}.
\textbf{}
\begin{figure}[h!]
	\centering
	\includegraphics{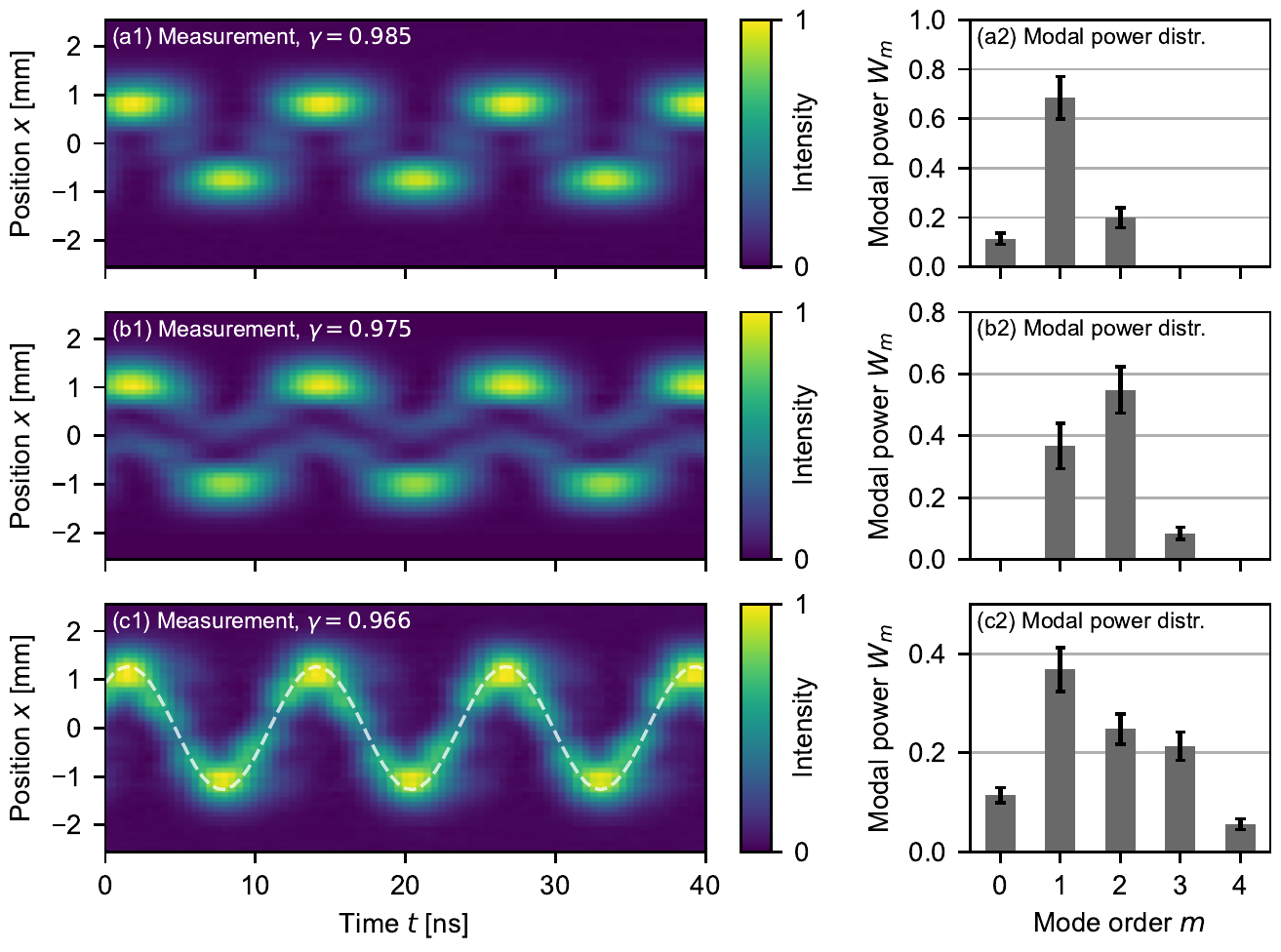}
	\caption{Changed spatio-temporal oscillations as a result of (a) a change in resonator length, leading to a set of superposed modes around $m_0=1$, (b) a reduced spatial offset $\Delta x = \SI{400}{\um}$, and (c) a RF modulation signal including harmonics at $\nu_0 \pm 2\Delta\nu$. The corresponding modal power distributions are shown on the right. The correlation coefficient $\gamma$ between each measured and numerically reconstructed spatio-temporal trajectory (shown in Supplement~1) is given. In (c1) the cosine trajectory expected for an ideal Poissonian power distribution with $\bar{m}=\num{2}$ is shown as dashed line. Intensity and modal powers are given in arbitrary units normalized to maximum intensity and to their overall sum, respectively.}
	\label{fig:result2}
\end{figure}

First, a superposition of a mode set around the central mode order $m_0=1$ was realized (see Fig.~\ref{fig:result2}(a)), whereas in the previously shown example only those instants -- while scanning the resonator length -- were considered, when the modes around $m_0=2$ were resonant.
Again, a spatio-temporal oscillation was observed, which agreed well with the reconstruction based on the measured modal powers with $\gamma=\num{0.985}$. 
The amplitude of the spatial oscillation, defined as the distance between center and turning points, decreased from \SI{0.93}{\mm} to \SI{0.73}{\mm}, because the size of a transverse mode is proportional to its mode order (approximately $\propto \sqrt{m}$)~\cite{Siegman1986}.
Hence, this superposition of modes with lower mode orders resulted in an oscillation with decreased size, i.e., spatial amplitude.
But most importantly, the chosen resonator length relative to the incident carrier frequency determined the central mode order $m_0$ and therefore allowed to precisely select the set of superposed transverse modes.
This high level of control over the set of locked modes is achieved by addressing the transverse modes by their spectral features, i.e., distinct resonance frequencies, improving the level of control, compared to relying on their similar spatial intensity distributions alone.

Second, the horizontal offset $\Delta x$ of the incident beam with regard to the optical axis of the resonator was reduced from \SI{580}{\um} to \SI{400}{\um}, which in turn reduced the spatial overlap $B_m$ with higher-order transverse modes~\cite{Verdeyen1968}.
As a result, the powers of both the $\text{HG}_{2,0}$ and $\text{HG}_{3,0}$ mode were decreased (see Fig.~\ref{fig:result2}(b)), whereas the power of the $\text{HG}_{1,0}$ mode was increased, compared to the measurement depicted in Fig.~\ref{fig:result1}.
This shift of power to lower mode orders slightly lowered the oscillation amplitude from \SI{0.93}{\mm} to \SI{0.85}{\mm}. 
Similar to the previous measurement (see Fig.~\ref{fig:result1}), the numerical reconstruction showed a constant intensity in the middle of the oscillation, that was not observed in the measurement, because the photodetector was ac-coupled.
There was still good agreement between measurement and reconstruction with $\gamma=\num{0.975}$.

Third, an RF signal containing the second harmonic of the modulation frequency $\Delta\nu$ was used to not only generate sidebands at $\nu_0 \pm \Delta\nu$ but also at $\nu_0 \pm 2\Delta\nu$. 
This procedure accomplished the simultaneous excitation of five transverse modes (see Fig.~\ref{fig:result2}(c)). 
Compared to the superposition of three transverse modes (see Fig.~\ref{fig:result1}) the scanning behavior of the beam was more pronounced: the intensity was less strictly confined to the turning points of the scanning motion and more evenly distributed over the course of the oscillation.
The spatial shape of the beam during its oscillation is a direct result of the set of superposed modes and their powers, i.e., the modal power distribution.
If the modal powers were Poisson distributed, the transverse mode-locked beam would oscillate along a cosine-shaped trajectory with almost uniform intensity in the spatio-temporal representation~\cite{Auston1968a} (see Fig.~S1 in Supplement~1 and Visualization~1).
The broader modal power distribution of five transverse modes (see Fig.~\ref{fig:result2}(c)) was already more similar to a Poisson distribution, compared to the distribution of three modes (see Fig.~\ref{fig:result1}).
Therefore, the resulting spatio-temporal trajectory also more clearly resembled a uniform cosine, as shown by the increased correlation of $\gamma=\num{0.917}$ between the spatio-temporal trajectory of five superposed modes and the trajectory resulting from a Poissonian modal power distribution (with mean $\bar{m}=2$), compared to $\gamma=\num{0.713}$ in the case of three superposed modes.

These examples demonstrate the defined altering of the spatial shape of the beam during its oscillation. 
Changing the resonator length by fractions of the incident beam's wavelength enabled to precisely select the central mode order, and the spatial offset allowed to shift the modal powers between low and high mode orders.
However, the most influential parameters were the number and amplitudes of the initial spectral components.
In this particular example, increasing the number of spectral components allowed to generate a spatio-temporal oscillation similar to one expected for an ideal Poissonian modal power distribution.

In this work, we demonstrated the conversion of longitudinal to transverse mode-locked states and highlighted the individual contributing factors.
In future conversion experiments, a transverse mode-locked state with a true Poissonian modal power distribution can be achieved by consistently exploiting that the spatial coupling coefficients are Poisson distributed if the incident beam matches the fundamental resonator mode in size but is displaced in a transverse direction~\cite{Verdeyen1968}.

\subsection{Comparison to other transverse mode-locking schemes}
Other approaches to generate transverse mode-locked beams include laser-active resonators.
Which transverse mode-locked state is generated is controlled by making use of the different spatial intensity distributions of the individual transverse modes through control of spatially dependent gain and loss in the resonator.
To establish gain for many transverse modes, the pump spot is transversely displaced from the optical axis~\cite{Liang2013, Shen2018} or shaped such that the gain distribution is similar to the desired transverse modes' intensity distributions~\cite{Dukhovnyi1971, Schepers2020}. 
Similarly, spatially dependent losses limit which transverse modes can be excited to a certain range of mode orders, for example by inserting a slit into the resonator~\cite{Auston1968, Dukhovnyi1971, Schepers2020}.
Additionally, the mode-locked state is also influenced by the pump power~\cite{Zhang2021}, because the oscillation threshold of each mode depends on its spatial intensity distribution~\cite{Kubodera1979}.
However, it becomes increasingly difficult to generate transverse mode-locked states with controlled modal power distributions when only leveraging the modes' spatial features, because their intensity distributions have considerable overlap and even become more similar with increasing mode order.
Therefore, the set of locked transverse modes and their respective powers could not be controlled independently.

In contrast, the approach presented here allows to control which modes are locked and their according powers separately by not only utilizing the similar spatial features of the transverse modes but also their distinct resonance frequencies within the resonator.
Providing an input beam with spectral components that match transverse modes' resonance frequencies, allows to select a specific set of adjacent transverse modes.
Then, the modal power distribution can be shaped by altering the spatial overlap between incident beam and transverse resonator modes as well as by altering the amplitudes of the incident spectral components.
This two-step approach allows to select which transverse modes, and to what extent, are involved in the locking-process, and therefore also enables an increased level of control over the spatial shape of the scanning beam. 
Additionally, in this experimental realization the incident spectral amplitudes depend on an electronic control signal and can thus be changed rapidly with switching times mainly limited by the photon lifetime in the resonator, e.g., \SI{1}{\us} for a \SI{1}{\m} long resonator with a finesse of \num{1000}.
Furthermore, this approach uses a resonator without gain medium, which enables the generation of transverse modes and phase-locked states independent of any lasing threshold.

\section{Conclusion and outlook}
We proposed and demonstrated the generation of transverse mode-locked states via conversion of incident longitudinal mode-locked states by means of an empty optical resonator. 
The spectral components of an incident longitudinal phase-locked beam are transmitted simultaneously by matching their frequencies to the resonator's transverse mode resonances, resulting in a superposition of phase-locked transverse modes.
As a result, an amplitude-modulated beam is converted into a beam that rapidly moves across the transverse plane, highlighting the connection between the temporal and the spatio-temporal domain formed by a resonator and its transverse eigenmodes.
The conversion scheme was verified experimentally as multiple mode-locked states with up to five transverse modes were generated in a controlled fashion by precisely altering the central mode order, the spatial overlap with the transverse resonator modes, and the amplitudes of the incident spectral components.
Utilizing these spatial and spectral features of transverse modes enables to control the set of locked modes independent of their powers, which gives rise to a new degree of freedom in the generation of transverse mode-locked states, compared to their excitation and locking in laser-active resonators.

In principle, the presented conversion scheme can also be reversed and should result in the conversion of a transverse mode-locked input, i.e., a scanning beam, to a longitudinal mode-locked output, i.e., a temporally pulsed beam.
This transfer of oscillations from the temporal to the spatio-temporal domain and vice versa may find applications in optical metrology.
Converting temporal to spatio-temporal oscillations enables measuring temporal phenomena in the spatial domain, for example by evaluating the resulting time-averaged spatial intensity distributions.
With the reversed conversion scheme, spatial oscillations are converted to temporal ones, giving access to information about spatial processes in the time domain without requiring spatially resolved detection.
Furthermore, the conversion scheme can also be applied to a longitudinal mode-locked laser beam, thereby imprinting a rapid scanning motion onto the beam.
The resulting temporally pulsed and spatially scanning beam is thus generated without any moving components or high-frequency electronic control signals, enabling purely passive high-speed beam scanning.

\begin{backmatter}
\bmsection{Funding}
Open Access Publication Fund of the University of Münster.

\bmsection{Acknowledgments}
We acknowledge support from the Open Access Publication Fund of the University of Münster.

\bmsection{Disclosures}
The authors declare no conflicts of interest.

\bmsection{Data Availability Statement}
Data underlying the results presented in this paper are not publicly available at this time but may be obtained from the authors upon reasonable request.

\bmsection{Supplemental document}
See Supplement~1 for supporting content.
\end{backmatter}


\bibliography{bib}

\end{document}


\maketitle

\section{Ideal spatio-temporal oscillation}
\begin{figure}[h]
	\centering
	\includegraphics{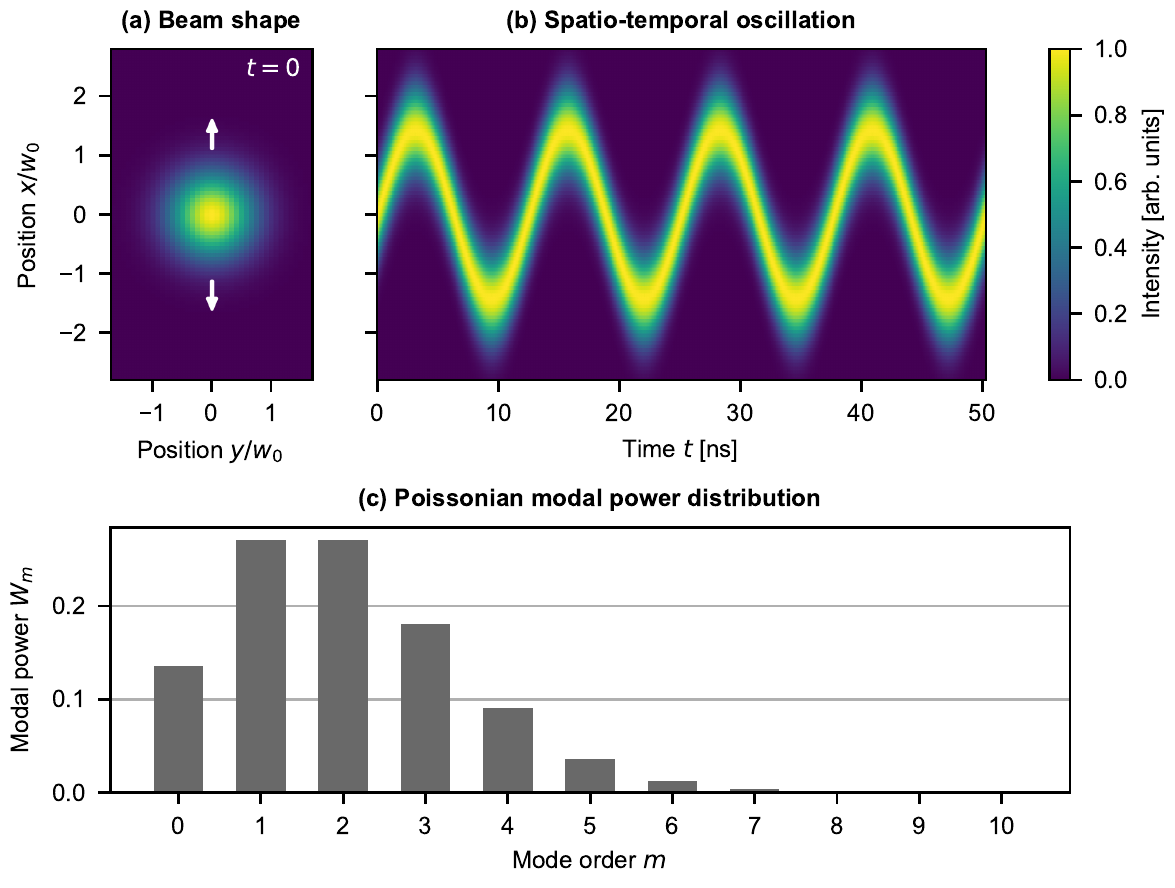}
	\caption{A transverse mode-locked state results in (a) a Gaussian spot that (b) oscillates across the transverse plane due to different transverse mode resonance frequencies, given that (c) the modes' powers are distributed according to a Poissonian (here with mean $\bar{m}=2$). The transverse coordinates $x$ and $y$ are normalized to the radius $w_0$ of the fundamental mode. A frequency difference of \SI{79.6}{\MHz} between adjacent transverse modes was assumed in accordance with the experimental setup presented in the primary document.}
	\label{fig:tml_ideal}
\end{figure}

The spatio-temporal oscillation originating from a superposition of Hermite-Gaussian $\text{HG}_{m,0}$ modes with a Poissonian modal power distribution is depicted in Fig.~\ref{fig:tml_ideal}. An animation of the spatio-temporal oscillation is provided as Visualization~1.

\section{Experimental results - longitudinal to transverse mode-locked state conversion}
In the following, multiple transverse mode-locked (TML) states are depicted that have been realized experimentally by conversion from incident longitudinal mode-locked states.
For each TML state the modal power distribution $W_m$ was determined, i.e., which modes are superposed and what are their respective powers (see Sec.~3 in the primary document). 
This modal power distribution then allowed to compute a numeric reconstruction of the spatio-temporal trajectory (see Subsec.~4.1 in the primary document), which was compared to the according measurement by the correlation coefficient $\gamma$ (see Eq.~5 in the primary document) for each TML state.
In contrast to the measured modal powers, the modal phases assumed for the reconstructions were derived from the architecture of the used amplitude modulator given the applied modulation signals.

Subsec.~\ref{subsec:1_initial} shows a listing of transverse mode-locked states resulting from the conversion of three incident longitudinal modes. Then, the spatial overlap between the incident beam and the resonator modes was reduced, resulting in the TML states depicted in Subsec.~\ref{subsec:2_reduced_spatial_overlap}. Subsec.~\ref{subsec:3_5_modes} features TML states consisting of five transverse modes. Finally, it is shown that non-linear modal phases led to different periodic spatio-temporal patterns, which can be seen in Subsec.~\ref{subsec:4_5_modes_nonlinear_phase}.

\subsection{Example 1 - initial measurement} \label{subsec:1_initial}
\begin{figure}[H]
	\centering
	\includegraphics{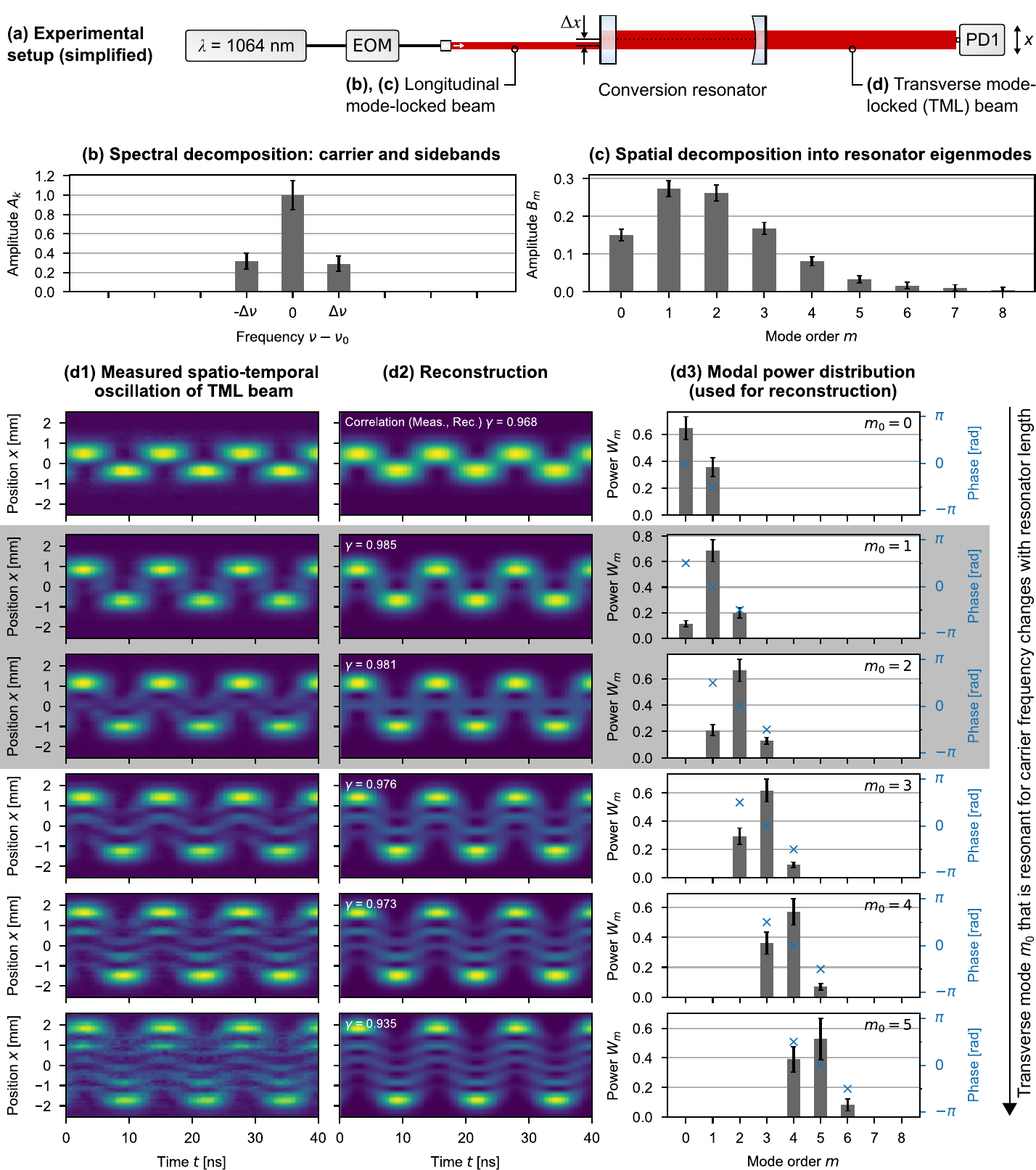}
	\caption{(a) An empty optical resonator is used to convert a longitudinal mode-locked state, characterized by its (b) different frequency components and (c) its beam shape with respect to the resonator eigenmodes, into (d) various transverse mode-locked states. The false-color plots in (d1) and (d2) depict spatio-temporal intensity distributions $I(t,x)$, each normalized to their respective maximum value. (d3) shows the set of superposed modes with their powers $W_m$ (normalized to their sum) and phases that were used for the reconstruction. The shaded measurements are also included in the primary document as Fig.~4(a) and Fig.~3, respectively. Abbreviations in the schematic: electro-optic amplitude modulator (EOM), photodetector (PD1).}
	\label{fig:1_initial}
\end{figure}

Initially, a longitudinal mode-locked state with three frequency components was generated by amplitude modulation. 
The horizontal displacement $\Delta x = \SI{580}{\um}$ between incident beam and resonator axis established spatial overlap with higher-order transverse resonator modes.
The central mode order $m_0$ that is resonant for the carrier frequency $\nu_0$ was changed by altering the resonator length by a fraction of the incident beam's wavelength, leading to superpositions of different transverse modes depicted in the rows of Fig.~\ref{fig:1_initial}(d).
The presented conversion scheme is validated by the good agreement between measured and reconstructed spatio-temporal trajectories with correlation coefficients $\gamma > 0.96$ for all but one mode-locked state, where noise due to low signal levels reduced the correlation to $\gamma=\num{0.935}$. 

\cleardoublepage

\subsection{Example 2 - reduced spatial overlap}\label{subsec:2_reduced_spatial_overlap}
\begin{figure}[H]
	\centering
	\includegraphics{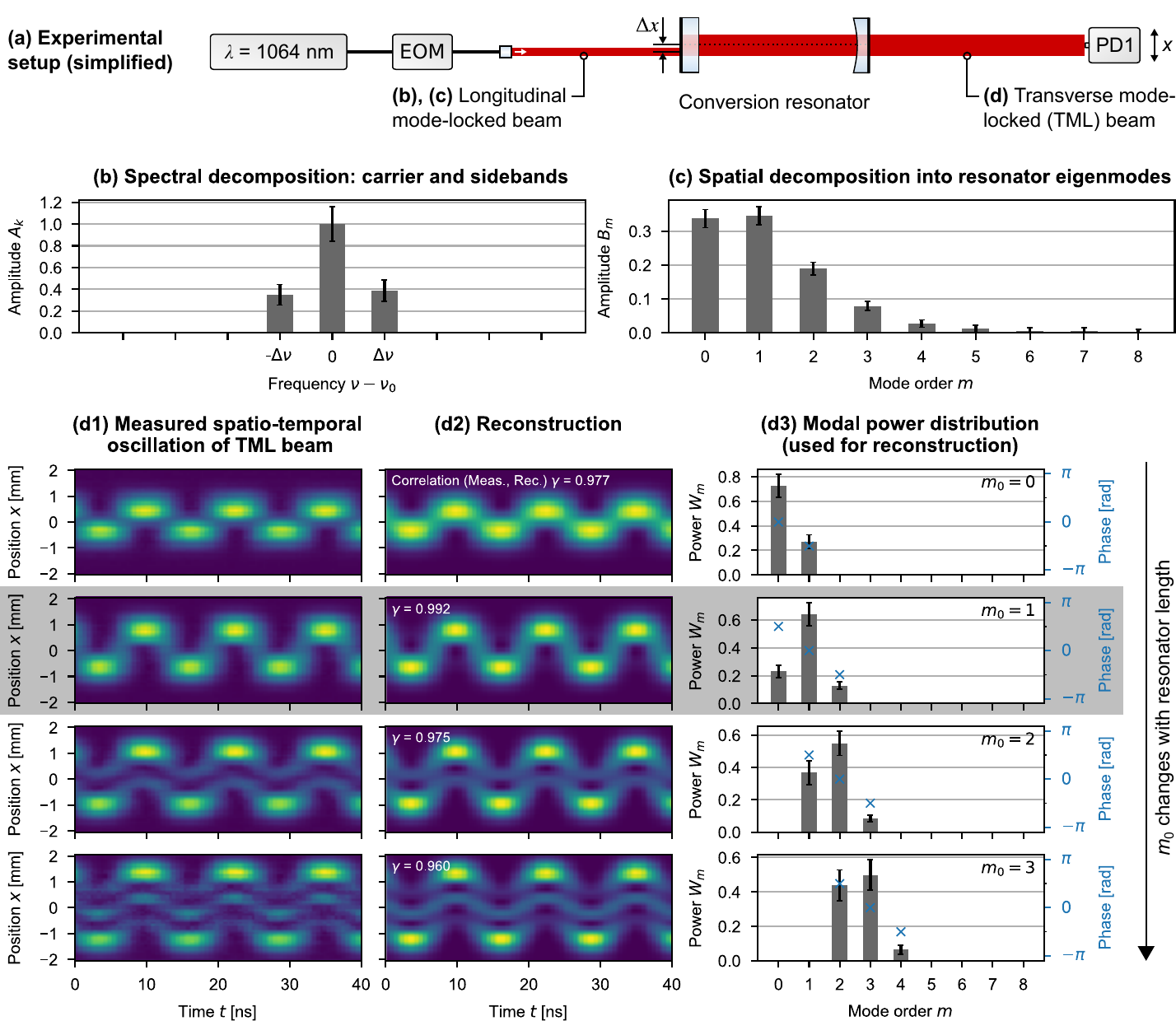}
	\caption{(a) An empty optical resonator is used to convert a longitudinal mode-locked state, characterized by its (b) different frequency components and (c) its beam shape with respect to the resonator eigenmodes, into (d) various transverse mode-locked states. The false-color plots in (d1) and (d2) depict spatio-temporal intensity distributions $I(t,x)$, each normalized to their respective maximum value. (d3) shows the set of superposed modes with their powers $W_m$ (normalized to their sum) and phases that were used for the reconstruction. The shaded measurement is also included in the primary document as Fig.~4(b). Abbreviations in the schematic: electro-optic amplitude modulator (EOM), photodetector (PD1).}
	\label{fig:2_reduced_spatial_overlap}
\end{figure}

Compared to the measurements depicted in Fig.~\ref{fig:1_initial} the spatial displacement $\Delta x$ of the incident beam was reduced to \SI{400}{\um}, leading to a decrease in spatial overlap with higher-order transverse modes (see Fig.~\ref{fig:2_reduced_spatial_overlap}(c)).

\cleardoublepage

\subsection{Example 3 - additional spectral components}\label{subsec:3_5_modes}
\begin{figure}[H]
	\centering
	\includegraphics{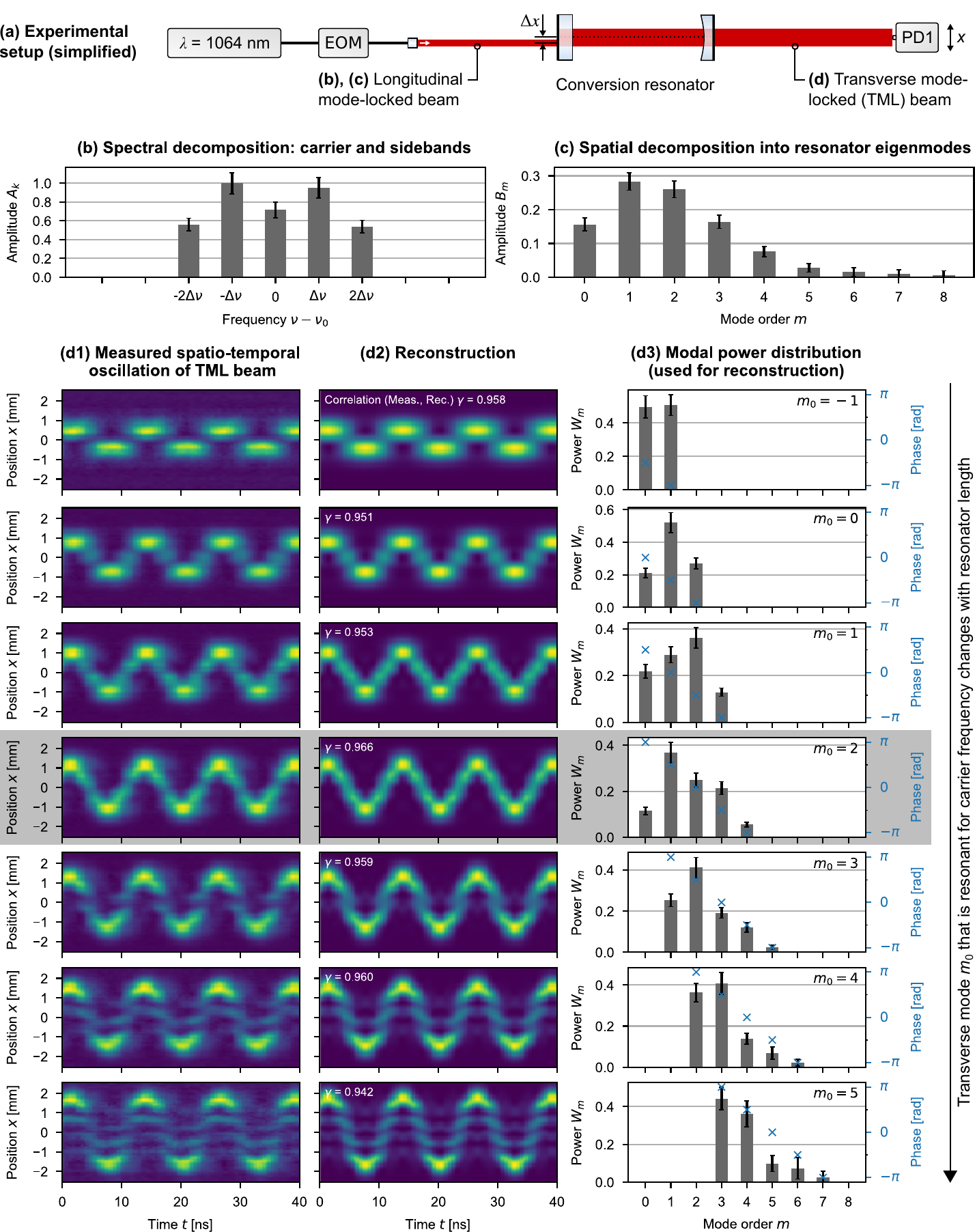}
	\caption{(a) An empty optical resonator is used to convert a longitudinal mode-locked state, characterized by its (b) different frequency components and (c) its beam shape with respect to the resonator eigenmodes, into (d) various transverse mode-locked states. The false-color plots in (d1) and (d2) depict spatio-temporal intensity distributions $I(t,x)$, each normalized to their respective maximum value. (d3) shows the set of superposed modes with their powers $W_m$ (normalized to their sum) and phases used for the reconstruction. The shaded measurement is also included in the primary document as Fig.~4(c). Abbreviations in the schematic: electro-optic amplitude modulator (EOM), photodetector (PD1).}
	\label{fig:3_5_modes}
\end{figure}

Two additional sidebands, i.e., frequency components, were generated by modulating the incident beam not only at $\Delta \nu$ but also at $2\Delta\nu$ (see Fig.~\ref{fig:3_5_modes}(b)).
A linear phase between all spectral components was ensured by setting an appropriate phase between the fundamental and the second harmonic modulation signal.

\cleardoublepage

\subsection{Example 4 - non-linear modal phases}\label{subsec:4_5_modes_nonlinear_phase}
\begin{figure}[H]
	\centering
	\includegraphics{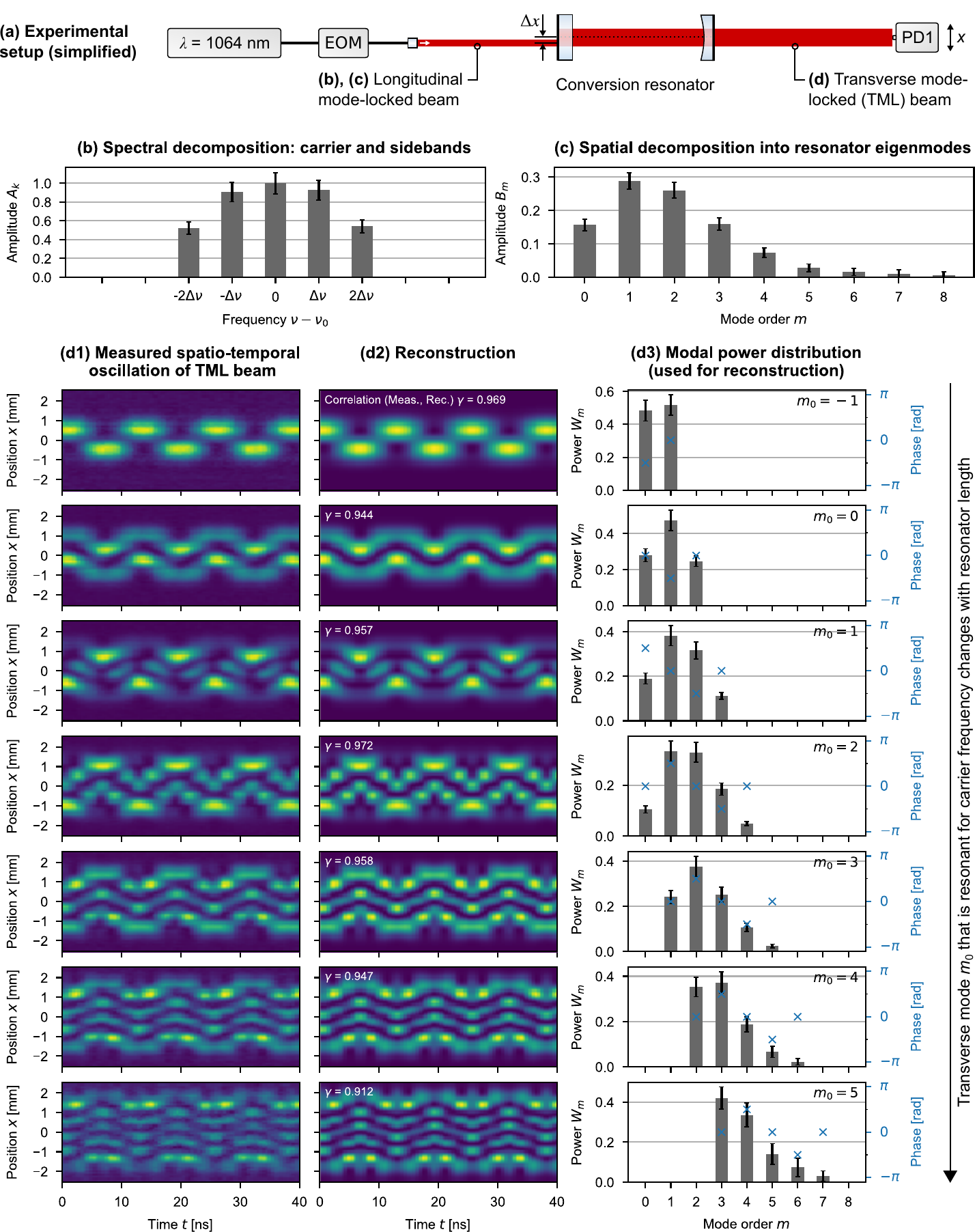}
	\caption{(a) An empty optical resonator is used to convert a longitudinal mode-locked state, characterized by its (b) different frequency components and (c) its beam shape with respect to the resonator eigenmodes, into (d) various transverse mode-locked states. The false-color plots in (d1) and (d2) depict spatio-temporal intensity distributions $I(t,x)$, each normalized to their respective maximum value. (d3) shows the set of superposed modes with their powers $W_m$ (normalized to their sum) and phases used for the reconstruction. Abbreviations in the schematic: electro-optic amplitude modulator (EOM), photodetector (PD1).}
	\label{fig:4_5_modes_nonlinear_phase}
\end{figure}

Again, a longitudinal mode-locked state consisting of five spectral components was generated.
However, the phases of the $\Delta\nu$ and the $2\Delta\nu$ modulation signals were deliberately chosen to result in non-linear modal phases in the transverse mode-locked output (see Fig.~\ref{fig:4_5_modes_nonlinear_phase}(d3)).
The resulting spatio-temporal patterns were still periodic, but showed multiple points of high intensity instead of forming single Gaussian-like spots oscillating across the transverse plane (see Fig.~\ref{fig:1_initial}, Fig.~\ref{fig:2_reduced_spatial_overlap}, Fig.~\ref{fig:3_5_modes}).

